\documentclass[preprint,12pt,authoryear]{elsarticle}
\usepackage{algorithm}
\usepackage{algpseudocode}
\usepackage{natbib}
\usepackage[left=2cm, right=2cm]{geometry}

\usepackage{amssymb}
\usepackage{bbm}
\usepackage{amsmath}
\usepackage{url}

\journal{Journal of Applied Statistics}

\begin{document}

\begin{frontmatter}



\title{Efficient Bayesian variable selection with reversible jump MCMC in imaging genetics: an application to schizophrenia}

 \author[inst1]{Djidenou Montcho}
 \affiliation[inst1]{organization={Statistics Program, CEMSE, King Abdullah University of Science and Technology},
             city={Thuwal},
             postcode={23955600},
             country={Kingdom of Saudi Arabia}}
 \author[inst2]{Daiane A. Zuanetti}
 \affiliation[inst2]{organization={Departamento de Estatistica, Universidade Federal de Sao Carlos},
             city={Sao Carlos},
             postcode={13565905},
             state={Sao Paulo},
             country={Brazil}}
 \author[inst3]{Thierry Chekouo}
 \affiliation[inst3]{organization={Division of Biostatistics, University of Minnesota},
             city={Minneapolis},
             postcode={55455},
             state={Minnesota},
             country={USA}}

 \author[inst2]{Luis A. Milan}
 
\begin{abstract}
From a practical perspective, proposals  are one of the main bottleneck for any  Markov Chain Monte Carlo (MCMC) algorithm. This paper suggests a  novel data driven  or informed  proposal for reversible jump MCMC for Bayesian variable selection in the context of predictive risk assessment for schizophrenia based on imaging genetic data.  Given functional Magnetic Resonance Image and Single Nucleotide Polymorphisms information of healthy and people diagnosed with schizophrenia, we use a Bayesian probit model to select discriminating variables for inferential purposes, while to estimate the predictive risk, the most promising models are combined using a Bayesian  model averaging scheme.
\end{abstract}




\end{frontmatter}




\section{Introduction}
\label{sec1}
Increasing computational power has enabled researchers to collect data of different types and sources, but
has also intensified the discussion on how to select a subset of variables with the best predictive performance or to better explain the phenomenon under analysis from an inferential or scientific perspective. In the Bayesian framework, model selection can be performed using a variety of techniques which are reviewed and compared in \cite{o2009review, gelman2014understanding}. Most used strategies could be classified into information criteria \citep{spiegelhalter2002Bayesian, chen2008extended, watanabe2010asymptotic},  Bayes factor \citep{kass1995Bayes}, shrinkage prior \citep{mitchell1988Bayesian,ishwaran2005spike, van2019shrinkage}, cross validation \citep{vehtari2017practical, liu2022leave}, and transdimensional algorithms such as stochastic search variable selection \citep{george1997approaches} and  reversible jump  Markov chain Monte Carlo (RJ) \citep{green1995reversible}, the focus of this work. 

The RJ algorithm  allows for a full Bayesian analysis, provides marginal posterior probability of inclusion for any covariate along with the posterior probability of visited models which could be combined for prediction using a  Bayesian model averaging scheme \citep{hoeting1999Bayesian}. However, it is not yet widely used because of the difficulty of its implementation, bad mixing, slow convergence due to a lack of straight strategy to design efficient proposals for inter and intra models moves. Usually, models are proposed based on the uniform distribution which is not our best option if the model space is very large, for example when selecting covariates from a large set, while candidates and parameters are sampled from some vague Gaussian or uniform distribution.  Furthermore, including information about the target distribution could increase the efficiency of MCMC (Markov chain Monte Carlo) when compared with methods based on naive, uniform or random walk. For instance, this is done in Hamiltonian Monte Carlo \citep{hmcneal} and Metropolis adjusted Langevin dynamics \citep{welling2011bayesian} using information from the gradient of the joint distribution. In the special context of RJ, many works have been dedicated to try to overcome these limitations \citep{brooks2003efficient, jain2004split, lamnisos2009transdimensional, saraiva2012clustering}. Recently, \cite{zanella2020informed} proposed locally balanced proposals for discrete spaces on top of which \cite{gagnon2019informed} also creates another informed RJ. A special informed RJ strategy proposed in \cite{zuanetti2016data} and also used in \cite{zuanetti2020bayesian}, named DDRJ (data driven reversible jump), makes use of the data to inform about the next best candidate model and has been proposed for mapping QTLs (Quantitative Trait Locus), i.e., selecting relevant genetic categorical covariates, which regulate quantitative traits. This methodology leads to a better mixing, improves the chain dynamic and effective sample size. 

In this work, our main contribution is that we build on top of the DDRJ and extends it to the context where we have categorical, numerical or both categorical and numerical covariates.  We propose a Bayesian predictive risk model for a binary variable, in particular the presence or absence of schizophrenia, based on a model averaging \citep{hoeting1999Bayesian} with sparse sets of neuroimaging and genetic covariates (imaging genetics) selected using an informed or data driven RJ. In addition, as the DDRJ provides the posterior probability of each model, we also combine the most visited models, using Bayesian model averaging, to create a classifier for future individuals and we compare its performance in terms of misclassification error and area under the receiver operating characteristic curve to our benchmark results in \cite{chekouo2016}, LASSO \citep{tibshirani1996regression} and random forest \citep{breiman1984classification}.

From the motivating problem, we have available fMRI (functional Magnetic Resonance Imaging) and SNP (Single Nucleotide Polymorphism) information on healthy and patients diagnosed with schizophrenia. fMRI was mainly designed to identify brain's response to task by detecting regional neuronal activity captured by blood oxygenation level-dependent (BOLD) variations. Actually, it is at the core of neuroimaging for studying schizophrenia because of its low invasiveness, absence of radiation and relatively high resolution. SNPs are substitutions of a single nucleotide at a specific position in the genome that occur in at least $1\%$ of the population. They are frequently used in Genome Wide Association Studies (GWAS) to find possible associations to disease and phenotypes \citep{mah2007gentle}. \cite{chen2012multifaceted} used principal and independent component analysis and found evidence of relevant association between fMRI and SNPs.  \cite{stingo2013integrative} extended this inferential problem and developed an integrative Bayesian hierarchical mixture model and applied it to link brain connectivity, through fMRI, to genetic information from SNPs of healthy and schizophrenic patients. \cite{chekouo2016} developed a Bayesian predictive model that includes ROIs (regions of interest) based network and a new network capturing relations between SNPs and ROIs to quantify a subject's risk of being schizophrenic based on fMRI and SNPs information. Auxiliary indicator variables with spike-slab priors (which may not be computationally scalable for large data sets) and a Bayesian model averaging were used for model selection and prediction, respectively.

This manuscript is organized as follow: Section \ref{modelo} proposes the Bayesian model under consideration to jointly select ROIs (numerical variables) and SNPs (categorical variables). The DDRJ algorithm and variable selection and prediction procedures are presented in Section \ref{algoritmo}. Section \ref{simulacao} shows their efficiency on simulated data and comparison with other selection and prediction methods. Finally, Sections \ref{mcic} and \ref{conclusao} contain the application of the methodologies to the Mind Clinical Imaging Consortium (MCIC) dataset and a discussion on results, and final considerations, respectively. The R codes and dataset used for implementing the methodologies are openly available in a public repository on Github at \url{https://github.com/hansamos/DDRJ}.

\section{Models for dichotomous traits} \label{modelo}

Given $n$ independent individuals, let $ \boldsymbol{Y}= (Y_1, \dots, Y_n) $ be the set of binary random variables, here characterizing their disease status, healthy or diagnosed with schizophrenia. Also consider the sets of covariates $\boldsymbol{X}=[X_{ip}]_{n \times g}$ and $\boldsymbol{Z}=[Z_{ik}]_{n \times m}$ as the matrices of $g$ numerical covariates (ROI-based summaries of blood oxygenation level-dependent, BOLD, intensity in this study) and $m$ categorical variables (genotype of SNPs in this study), respectively for $n$ subjects. 

To model the probability of success (suffering from schizophrenia in this study), we consider the probit data augmentation \citep{albert1993Bayesian} that introduces a continuous non-observable latent random variable  $Y_{i}^{*}$, normally distributed, and classifies the individual output according to its value being above a threshold or not. The variable $Y_{i}^{*}$ is viewed as a hidden process that depends on numerical and categorical covariates (ROIs and SNPs), such that when its value is positive, the individual is classified as a success (schizophrenic) and a failure (healthy) otherwise. Assuming the probit model leads us to well known conditional distributions for parameters and latent variables and allows to use Gibbs sampling for intra model updates. An alternative would be to use the data augmentation model proposed in \cite{polson2013Bayesian} for a Bayesian logit model, but this requires adding and updating P{\'o}lya-Gamma variables to obtain a simpler and more efficient simulation algorithm. 

Then, latent variable $Y_{i}^{*}$ is defined as
\begin{equation}
 Y_{i}^{*} =  \beta_0 +  \sum_{p \in\ \mathcal{G}}\beta_p X_{ip} + \sum_{k \in \mathcal{M}} \alpha_k Z_{ik} + \sum_{k \in \mathcal{M}} \delta_k (1-|Z_{ik}|) + \xi,\;    \xi \overset{iid}{\sim} N(0,1), 
 \label{model_geral}
\end{equation}
and $Y_i =\mathbbm{1}(Y^{*}_{i} >0)$ where $\mathbbm{1}(.)$ is the indicator function, $Z_{ik} \in \{-1,0,1 \}$, for SNPs having genotype aa, aA and AA, respectively. One could have simply used dummy variables to encode the categorical variables, such as the genotype of the SNPs. However, in Biology, the genetic interpretation is meaningful when the SNPs are encoded as we have done above \citep{zuanetti2016data}. The sets $\mathcal{G} $ and $\mathcal{M}$ contain the numerical (ROI) and categorical (SNP) covariates indices, respectively, present in a given model. More specifically, $\beta_p = 0 \;\text{if}\; p \notin \mathcal{G},\;  \alpha_k =\delta_k = 0  \;\text{if}\; k \notin \mathcal{M}$ and $\beta_p, \alpha_k, \delta_k$ are non zero, otherwise. Regarding the coefficients, $\beta_0$ is the intercept, $\beta_p$ is the effect of numerical covariate (ROI) $p$ while $\alpha_k$ and $\delta_k$ account for the additive and dominant effects of SNP $k$ for every $k=1,\dots,m$, respectively. 

Our goal is to select, under a Bayesian framework, a set of discriminatory  (ROIs) and  (SNPs) covariates from the set of available $g$ numerical (ROIs) and $m$ categorical (SNPs) covariates, respectively. We also aim at providing estimates for the coefficients $\beta_0$, $\beta_p$ and for the additive and dominant effects $\alpha_k$, $\delta_k$, respectively, for the selected features and identifying how they regulate and impact the chance of the success (schizophrenia). In addition, we intend to have a model with good predictive capacity as well.

Let us denote the unknown parameters by $\boldsymbol{\theta}=(\boldsymbol{\gamma},K,P)$ with $\boldsymbol{\beta}^{T}=(\beta_0, \beta_1, \dots, \beta_P)$, $ \boldsymbol{\alpha}^{T}=(\alpha_1, \dots, \alpha_K)$, $ \boldsymbol{\delta}^{T}=(\delta_1, \dots, \delta_K)$,  $\boldsymbol{\gamma}^{T}=(\boldsymbol{\beta}^{T}, \boldsymbol{\alpha}^{T}, \boldsymbol{\delta}^{T})$, $K= |\mathcal{M}|$  and $P= |\mathcal{G}|$. The likelihood function for $\boldsymbol{\theta}$ is given by
\begin{align}
\label{likelihood}
L(\boldsymbol{\theta}| \boldsymbol{Y},\boldsymbol{Y}^*,\boldsymbol{X,Z}) 
&=  \prod_{i=1}^{n} P(Y_{i}|Y_{i}^{*})P(Y_{i}^{*}\boldsymbol{|\theta, X,Z})  \nonumber \\ 
&= \frac{1}{(\sqrt{2\pi} )^n } \exp\left[-\frac{1}{2}\sum_{i=1}^{n} \xi_{i}^{2}\right] \nonumber \\ 
& \times \prod_{i=1}^{n}[\mathbbm{1}(Y_{i}=0)\mathbbm{1}(Y^{*}_{i}<0)+\mathbbm{1}(Y_{i}=1)\mathbbm{1}(Y^{*}_{i} \geq 0)],
\end{align} where
\begin{align}
\xi_i = Y^{*}_{i} -
     \beta_0 - \sum_{p \in \mathcal{G}}\beta_p X_{ip} - \sum_{k \in \mathcal{M}} \alpha_k Z_{ik} - \sum_{k \in \mathcal{M}} \delta_k (1-|Z_{ik}|).  \nonumber
\end{align}
 
We complete the model assigning independent prior distribution to each parameter and the joint prior distribution is defined by 
 \begin{equation}  \pi(\boldsymbol{\theta})=\pi(K)\pi(P) \pi(\boldsymbol{\beta}|P)\pi(\boldsymbol{\alpha}|K)\pi(\boldsymbol{\delta}|K),
     \label{prior modelo geral}
\end{equation}     
where, we assume that,
\begin{align}
K  \sim   \text{Unif}(m), P \sim   \text{Unif}(g),  
\boldsymbol{\beta} \sim  N_{P+1}(\mathbf{0},\sigma^{2}_{ \boldsymbol{\beta}}\mathbf{I}_{P+1} ),  \nonumber \\
\boldsymbol{\alpha}  \sim  N_{K}(\mathbf{0},\sigma^{2}_{ \boldsymbol{\alpha}}\mathbf{I}_{K} ),  
\boldsymbol{\delta}  \sim  N_{K}(\mathbf{0},\sigma^{2}_{ \boldsymbol{\delta}}\mathbf{I}_{K} )  
\end{align}
 with all hyperparameters $\sigma^{2}_{ \boldsymbol{\beta}},\;  \sigma^{2}_{ \boldsymbol{\alpha}},\;  \sigma^{2}_{ \boldsymbol{\delta}}$ fixed and $\mathbf{I}_d$ represents an identity matrix of dimension $d$.

The model in Equation \eqref{model_geral} is a classical regression model with Gaussian priors for the coefficients. Hence, all full conditional posterior for the parameters are Gaussian  and given by  
\begin{align}
\boldsymbol{\beta}|. \sim N(\boldsymbol{\beta}^*, \Gamma_1), \; 
\boldsymbol{\alpha}|. \sim N(\boldsymbol{\alpha}^*, \Gamma_2), \;
\boldsymbol{\delta}|. \sim N(\boldsymbol{\delta}^*, \Gamma_3), \label{post__alpha_beta_delta}
\end{align}
and the full conditional for the latent variable is a truncated Normal (Nt) distribution given by
 \begin{align}
Y^{*}_{i}|Y_{i}=1, . \sim \text{Nt}(\tilde{y}^{*}_{i}, 1 ,\text{left}=0),\;
Y^{*}_{i}|Y_{i}=0, . \sim  \text{Nt}(\tilde{y}^{*}_{i}, 1 ,\text{right}=0) \label{postlatent}.
\end{align}

Given the full conditional posteriors, described in more details in Appendix A in the supplementary material, we use a Gibbs sampling procedure to update the parameters iteratively given $K$ and $P$, in intra-model move. In the next section, we describe the data driven reversible jump algorithm (DDRJ) to efficiently propose the inter models move, where the candidate model consists of a previous model with the inclusion (birth) or removal (death) of a covariate to update $K$ or $P$.

\section{Data driven reversible jump for updating $K$ and $P$} \label{algoritmo}
Despite its generalization, RJ's performance relies on the probability of visiting the next model and the proposal distribution to obtain the next set of parameters within each model. Indeed, bad proposals will usually lead to high rejection rate, slow mixing and consequently more iterations would be needed for convergence. 
One reason to understand these points is that there is a high probability of rejecting a move from a parameter set with high density in a bad model to a parameter set with low density in a good model. And if the proposals are bad, this kind of move may be frequent and not accepted. 

Our proposal then, to select variables, is try to include or exclude a single covariate from the current model in a more efficient way. Thus, first, we decide if we will include a new covariate (birth) or exclude (death) one that is present in the current model. Obviously, in the case where we do not have any covariate in the model, i.e., a model with an intercept only, we would opt for a birth move with probability $1$ and, at the other extreme, when the model is saturated with all the possible covariates $(m+g)$, we would opt for a death move with probability $1$. After that, we define a measure roughly understood as a criterion to choose the next candidate, i.e.,  the covariate that should be excluded or included to the current model. After obtaining the candidate model, we sample the set of parameters for it and test its acceptance. 

As we have both numerical (ROIs) and categorical variables (SNPs) to be selected in an integrative manner, i.e., jointly, we could think of three alternatives to perform this joint variable selection. As the first option, we could select all possible numerical (ROIs) and then select categorical (SNPs) covariates, i.e., run the method considering only ROIs, then run the method for selecting SNPs conditional on selected ROIs. As a second option, we could select all possible SNPs and then select ROIs conditional on these selected SNPs. The last option is to randomly alternate between selecting numerical and categorical variables. Options 1 and 2 are special cases of the last option, thus we focus on describing how to carry the third option. However, we highlight that options 1 and 2 may be computationally more efficient and show better convergence when dealing with very high-dimensional data.

More importantly, instead of using a uniform distribution to choose which categorical (SNP) or numerical covariate (ROI) will be included or excluded from the model, we prioritize those covariates that seem to be more or less associated with the trait conditioned on the current model. For measuring the covariates association with the trait conditioned on the current model, we use  different measures for categorical and numerical covariates. For the numerical covariates (ROIs), we use the Pearson correlation coefficient between each covariate and the residuals of the current model, while for categorical covariate (SNPs), we use the Kruskal-Wallis (KW) statistics between each variable and the residuals of the current model. One could choose a different criterion to measure the quality of a candidate, and from our experiment the efficiency of the DDRJ also depends on that. The KW measure  was used by \cite{zuanetti2016data} for QTL mapping with categorical covariates and continuous trait, thus our innovation here for inferential goals is to use the KW and Pearson correlation for jointly selecting categorical and numerical covariates considering binary trait. 

At each stage of the process, we randomly alternate between numerical and categorical covariates in the following manner. Decide with probability $s=\frac{g}{m+g}$ and $1-s=\frac{m}{m+g}$ to work on ROIs or SNPs, respectively. This step allows us to jump into ROIs or SNPs space and then work on them separately.  This is fair if $ m \approx g$ as $ s \approx 0.5$. However, if one dimension dominates the other, it may be better to select variables separately or simply design an informed probability to favor any desired space. If numerical covariates space has been selected, then we apply the method described in Section \ref{ROIspace} conditional on already selected SNPs and ROIs up to this stage. If categorical space has been selected, then we apply the method described in Section \ref{SNPspace} conditional on already selected ROIs and SNPs at this moment. 

\subsection{Jumping into numerical covariates space}
\label{ROIspace}
Suppose that the current model contains $P= |\mathcal{G}|$  ROIs and $K =|\mathcal{M}|$  SNPs, with parameters $\boldsymbol{\theta}=(\boldsymbol{\beta}^{T},\boldsymbol{\alpha}^{T}, \boldsymbol{\delta}^T, K , P)$ and we decide to jump to ROIs (numerical) space. If $P=0$ then a birth ($b$) movement is proposed with probability $p(b|P=0)=1$,  when  $0<P<g$,  a birth or death movement is proposed with probability $p(b|P)=p(d|P)=\frac{1}{2}$ and finally  if $P=g$ then a death (d) movement is proposed with probability $p(d|P=g)=1.$

\begin{enumerate}
\item \textbf{Birth}:
Let's suppose that a birth move has been chosen. We propose to choose the next candidate from the remaining ROIs in $ \boldsymbol{X}_{-\mathcal{G}} = \{ X_p : p \notin \mathcal{G}\}$ with probability $p_{bj} = \frac{ |cor(\xi, X_j)|}{\sum_{X_p \in \boldsymbol{X}_{-\mathcal{G}}}|cor( \xi, X_p)|} \label{nasce_roi_modelo_3}$, where $cor(\xi, X_p) $ is the correlation between a candidate ROI $X_p$ and the residuals $\xi$ from the current model in Equation \eqref{model_geral}. Instead of uniformly choosing from the set of remaining ROIs, the main idea of our data driven proposal is to choose the  ROI which is highly correlated to the residuals of the current model. To speed up computation, one could use only part of the data to compute  $p_{bj}$.

After selecting a ROI $ X^b$, with index $b$, to be added to $\mathcal{G}$, we sample $\boldsymbol{\theta}^b$ from the conditional posterior distributions of $\boldsymbol{\beta}^b$, $\boldsymbol{\alpha}^b$ and $\boldsymbol{\delta}^b$ and test its acceptance with probability $\psi^b= min(1,A^b)$, where
\begin{equation}
    A^b= \frac{L( \boldsymbol{\theta}^{b}|\boldsymbol{X}_{\mathcal{G} \cup \{b\}}, \boldsymbol{Z}_{\mathcal{M}},\boldsymbol{Y}^* )\pi(\boldsymbol{\theta}^{b})q(\boldsymbol{\theta}|\boldsymbol{\theta}^b) }{L(\boldsymbol{\theta}|\boldsymbol{X}_{\mathcal{G}},\boldsymbol{Z}_{\mathcal{M}},\boldsymbol{Y}^* )\pi(\boldsymbol{\theta}) q(\boldsymbol{\theta}^{b}|\boldsymbol{\theta})},  \label{include_roi_model3}
\end{equation} 
$$q(\boldsymbol{\theta}^{b}|\boldsymbol{\theta})= p(b|P) p_{bj} \pi(\boldsymbol{\theta}^b|\boldsymbol{X}_{\mathcal{G} \cup \{b\}}, \boldsymbol{Z}_{\mathcal{M}},\boldsymbol{Y}^{*}),$$
$$q(\boldsymbol{\theta}|\boldsymbol{\theta}^b )= p(d|P+1)p_{dj}\pi(\boldsymbol{\theta}| \boldsymbol{X}_{\mathcal{G}}, \boldsymbol{Z}_{\mathcal{M}}, \boldsymbol{Y}^{*}),$$ $\boldsymbol{X}_{\mathcal{G} \cup \{b\}}$ of dimension $n \times (P+1)$ is the updated design matrix with the new ROI $X^b$ and $p_{dj}$ the probability of death (exclusion) that will be better explained in the death step.

The proposal distribution  $q(\boldsymbol{\theta}^{b}|\boldsymbol{\theta})= p(b|P) p_{bj}\pi(\boldsymbol{\theta}^b|\boldsymbol{X}_{\mathcal{G} \cup \{b\}}, \boldsymbol{Z}_{\mathcal{M}},\boldsymbol{Y}^{*})$ is a simple application of   conditional probabilities as  the new model and parameters are obtained from a sequence of  3 conditional steps. First,  we choose a  birth move  with probability $p(b|P)$,  then we choose the variable  to be included to obtain the new model with probability $p_{bj}$ and finally we sample the new parameters using the full conditional with probability $\pi(\boldsymbol{\theta}^b|\boldsymbol{X}_{\mathcal{G} \cup \{b\}}, \boldsymbol{Z}_{\mathcal{M}},\boldsymbol{Y}^{*})$. This idea applies regardless of birth or death movement.
\\
\item \textbf{Death}:
If on the other side, a death has been selected, then a possible way of choosing the candidate ROI to be deleted is by comparing the size of their coefficients after scaling the design matrix. Thus, we propose to select a ROI to be excluded with probability $p_{dj}= \frac{ \frac{1}{ |\beta_{j}| } }{ \sum_{p \in \mathcal{G}} \frac{1}{|\beta_{p}|}} \label{morte modelo 2}$. The larger the coefficient  of a given ROI, the smaller is its probability to be deleted from the current model.

After selecting a ROI $X^d$, with index $d$, to be deleted from $\mathcal{G}$, we sample $\boldsymbol{\theta}^d$ from the conditional posterior distributions of $\boldsymbol{\beta}^d$, $\boldsymbol{\alpha}^d$ and $\boldsymbol{\delta}^d$ and test its acceptance with probability $\psi^d= min(1,A^d)$, where 
\begin{equation}
    A^d= \frac{L(\boldsymbol{\theta}^{d}|\boldsymbol{X}_{\mathcal{G} \setminus \{d\}},\boldsymbol{Z}_{\mathcal{M}},\boldsymbol{Y}^* )\pi(\boldsymbol{\theta}^{d})q(\boldsymbol{\theta}|\boldsymbol{\theta}^{d}) }{L(\boldsymbol{\theta}|\boldsymbol{X}_{\mathcal{G}}, \boldsymbol{Z}_{\mathcal{M}},\boldsymbol{Y}^* )\pi(\boldsymbol{\theta})q(\boldsymbol{\theta}^{d}|\boldsymbol{\theta}) }, \label{exlude_roi_model3}
\end{equation}
$$q(\boldsymbol{\theta}^{d}|\boldsymbol{\theta})= p(d|P)p_{dj}\pi(\boldsymbol{\theta}^d|\boldsymbol{Y}^*,\boldsymbol{X}_{\mathcal{G} \setminus \{d\}}, \boldsymbol{Z}_{\mathcal{M}}),$$  $$q(\boldsymbol{\theta}|\boldsymbol{\theta}^d )= p(b|P-1)p_{bj}\pi(\boldsymbol{\theta}|\boldsymbol{Y}^*,\boldsymbol{X}_{\mathcal{G}},\boldsymbol{Z}_{\mathcal{M}}),$$ and $\boldsymbol{X}_{\mathcal{G} \setminus \{d\}}$ of dimension $n \times (P-1)$ is the updated design matrix without the deleted ROI $X^d$.
\end{enumerate}

\subsection{Jumping into categorical covariates space}
\label{SNPspace}
Under the same setting, suppose that the current model contains $P= |\mathcal{G}|$ ROIs and $K =|\mathcal{M}| $ SNPs, with parameters $\boldsymbol{\theta}=(\boldsymbol{\beta}^T,\boldsymbol{\alpha}^T, \boldsymbol{\delta}^T, K, P)$ and we decide to jump into SNPs (categorical covariates) space. In the same way as we did for ROIs, if $K=0$ a birth ($b$) movement is proposed with probability $p(b|K=0)=1$, if $0<K<m$ then a birth or death movement is proposed with probability $p(b|K)=p(d|K)=\frac{1}{2}$ and when $K=m$ then a death ($d$) movement is proposed with probability $p(d|K=m)=1.$

\begin{enumerate}
\item \textbf{Birth}:    
The choice of the next SNP to be included is guided by its association with the residuals $\xi$ from model in Equation \eqref{model_geral}. Each SNP $Z_{k}$ is a factor with 3 levels, so its  association with the current residuals can be measured using the Kruskal-Wallis (KW) statistics. Therefore $Z_{k}$  is selected from the set of remaining SNPs $ \boldsymbol{Z}_{-\mathcal{M}} = \{ Z_k : k \notin \mathcal{M}\}$  with probability $p_{bk}=\frac{\text{KW}(\xi, Z_{k}) }{\sum_{ Z_k\in \boldsymbol{Z}_{-\mathcal{M}}} \text{KW}(\xi, Z_{k})}
\label{nasce_snpmodelo3}$ . It's worth mentioning that we are not testing hypothesis but only using the test's statistic as a measure  to quantify levels of association.
    
After selecting a SNP $Z^b$, with index $b$, to be added to $\mathcal{M}$, we sample $\boldsymbol{\theta}^b$ from the conditional posterior distributions of $\boldsymbol{\alpha}^b$, $\boldsymbol{\delta}^b$ and $\boldsymbol{\beta}^b$ and test its acceptance with probability $\psi^b= min(1,A^b)$, where
\begin{equation}
    A^b= \frac{L( \boldsymbol{\theta}^{b}|\boldsymbol{X}_{\mathcal{G}}, \boldsymbol{Z}_{\mathcal{M} \cup \{b\}},\boldsymbol{Y}^* )\pi(\boldsymbol{\theta}^{b})q(\boldsymbol{\theta}|\boldsymbol{\theta}^b) }{L(\boldsymbol{\theta}|\boldsymbol{X}_{\mathcal{G}},\boldsymbol{Z}_{\mathcal{M}},\boldsymbol{Y}^* )\pi(\boldsymbol{\theta}) q(\boldsymbol{\theta}^{b}|\boldsymbol{\theta})}, \label{include_snp_modelo3}
\end{equation}
$$q(\boldsymbol{\theta}^{b}|\boldsymbol{\theta})= p(b|K) p_{bk} \pi(\boldsymbol{\theta}^b|\boldsymbol{X}_{\mathcal{G}}, \boldsymbol{Z}_{\mathcal{M} \cup \{b\}},\boldsymbol{Y}^{*}),$$ $$q(\boldsymbol{\theta}|\boldsymbol{\theta}^b )= p(d|K+1)p_{dk}\pi(\boldsymbol{\theta}| \boldsymbol{X}_{\mathcal{G}}, \boldsymbol{Z}_{\mathcal{M}}, \boldsymbol{Y}^{*}),$$ $\boldsymbol{Z}_{\mathcal{M} \cup \{b\}}$ of dimension $n \times (K+1)$ is the updated design matrix with the new SNP $Z^b$ and   $p_{dk}$ the probability of death (exclusion) defined in the death step.
\\
\item \textbf{Death}: As  $Z_{k}$ only takes value in $\{-1,0,1\}$, the absolute value of the coefficients $\boldsymbol{\alpha}_k$ and $\boldsymbol{\delta}_k$ in Equation \eqref{model_geral} give a measure of its importance. We propose to select a SNP to be excluded from the current model with probability $p_{dk}= \frac{\frac{1}{|\alpha_k|+|\delta_k|}  }{ \sum_{k \in \mathcal{M}} \frac{1}{|\alpha_k|+|\delta_k|}} \label{morte_snpmodelo3}$. The higher the effect of the SNP, the lesser is its probability to be deleted.

After selecting a SNP $Z^d$, with index $d$, to be excluded from $\mathcal{M}$, we sample $\boldsymbol{\theta}^d$ from conditional posterior distributions for $\boldsymbol{\alpha}^d$, $\boldsymbol{\delta}^d$ and $\boldsymbol{\beta}^d$ and test its acceptance with probability $\psi^d= min(1,A^d)$, where
\begin{equation}
    A^d= \frac{L(\boldsymbol{\theta}^{d}|\boldsymbol{X}_{\mathcal{G}},\boldsymbol{Z}_{\mathcal{M} \setminus \{d\}},\boldsymbol{Y}^* )\pi(\boldsymbol{\theta}^{d})q(\boldsymbol{\theta}|\boldsymbol{\theta}^{d}) }{L(\boldsymbol{\theta}|\boldsymbol{X}_{\mathcal{G}}, \boldsymbol{Z}_{\mathcal{M}},\boldsymbol{Y}^* )\pi(\boldsymbol{\theta})q(\boldsymbol{\theta}^{d}|\boldsymbol{\theta}) }, \label{exclude_snp_modelo3}
\end{equation}
$$q(\boldsymbol{\theta}^{d}|\boldsymbol{\theta})= p(d|K)p_{dk}\pi(\boldsymbol{\theta}^d|\boldsymbol{Y}^*,\boldsymbol{X}_{\mathcal{G}}, \boldsymbol{Z}_{\mathcal{M} \setminus \{d\}}),$$  $$q(\boldsymbol{\theta}|\boldsymbol{\theta}^d )= p(b|K-1)p_{bk}\pi(\boldsymbol{\theta}|\boldsymbol{Y}^*,\boldsymbol{X}_{\mathcal{G}},\boldsymbol{Z}_{\mathcal{M}}),$$ and $\boldsymbol{Z}_{\mathcal{M} \setminus \{d\}}$ of dimension $n \times (K-1)$ is the updated design matrix without the deleted SNP $Z^d$.
\end{enumerate}

The algorithm for performing joint selection for ROIs (numerical covariates) and SNPs (categorical covariates) and estimating the models' coefficients is summarized in Appendix B in the supplementary material.

\bigskip
Discussing about the validity of the DDRJ acceptance probabilities, consider a birth movement from a model $M$  to a model $M^b$ with parameters  $\boldsymbol{\theta}$ and $\boldsymbol{\theta}^{b}$ respectively. Let $\mathbf{u}=\boldsymbol{\theta}^{b}$ be the auxiliary variables of the transition $M\rightarrow M^{b}$, and $\mathbf{u}^{b}=\boldsymbol{\theta}$ be auxiliary variables of the transition $M^{b} \rightarrow M$ which represents a death movement.

In this way, the transition $M\rightarrow M^{b}$ involves a deterministic map function $h(\boldsymbol{\theta}, \mathbf{u})=(\mathbf{u}^{b},\boldsymbol{\theta}^{b})$ where the proposal density of $\mathbf{u}=\boldsymbol{\theta}^{b}$ is composed of the conditional posterior distributions used to simulate $\boldsymbol{\theta}^{b}$ and $h(\cdot,\cdot)$ is one-to-one function with unity Jacobian. Therefore, the proposed DDRJ method to update $K$ and $P$ is a special case of the traditional reversible jump algorithm and the proposed chain is ergodic and its convergence to the desirable invariant distribution is guaranteed.

\subsection{Variable selection and prediction procedures}
As stated at the beginning of the manuscript, our goal is to use the proposed method for variable selection and to carry out prediction for new individuals as well. For variable selection, the full dataset is used as training, which allows us to have a greater sample size to check the inferential performance of the method. We decide to select as relevant only those covariates with marginal posterior probability of inclusion (mppi), estimated as their relative frequency of being present in the  models, above a threshold ($0.5$ for instance). To assess the model's predictive performance, we use a 5-fold cross validation approach.
 

To predict the success (here the disease status) for a new individual having numerical (ROIs) and categorical (SNPs) covariates given by $\boldsymbol{X}^{new}$ and $\boldsymbol{Z}^{new}$, first we need to predict its non-observable variable $Y^{*}_{new}$ via a Bayesian model averaging as  
\begin{equation}
\hat{y}^{*}_{new}= \sum_{t} \left(\hat{\beta}_{0}^{t} + \sum_{p \in \mathcal{G}^t}\hat{\beta}_{p}^{t} X_{p}^{new} + \sum_{k \in \mathcal{M}^t} \hat{\alpha}_{k}^{t} Z_{k}^{new} + \sum_{k \in \mathcal{M}^t} \hat{\delta}_{k}^{t} (1-|Z_{k}^{new}|), \right)P(M_{t}|\boldsymbol{Y})
\end{equation}
where the index $t$ represents each of the $M_t$ models visited during the MCMC iterations, the parameters' estimates for each one are set to be their posterior mean and $P(M_{t}|\boldsymbol{Y})$ is the marginal posterior probability of the model $M_t$. Then, the posterior predictive probability of success (disease) for the new individual is computed as $P(Y_{new}=1|\Omega)= \Phi(\hat{y}^{*}_{new}|\Omega)$, where $\Phi(.)$ represents the standard normal cumulative distribution function and  $\Omega$ considers all parameters and data. If $P(Y_{new}=1|\Omega)= \Phi(\hat{y}^{*}_{new}|\Omega)>0.5$, the individual is classified as a success (schizophrenic). Here, instead of using the posterior predictive distribution of $Y_{new}$ as is usually done in Bayesian model averaging, we propose a point prediction defined as the weighted average of the models' predictions such as what is done by ensemble models and for computational ease.

From these posterior probabilities and non-observable variables, we can compute the AUC (area under the ROC curve) and MCE (misclassification error) to assess the predictive performance of the method in terms of variable selection and prediction, respectively.

\section{Simulation study}
\label{simulacao}
This section summarizes a simulation study to demonstrate the efficiency of the proposed method for performing variable selection using DDRJ and for making prediction for future individuals.  For each scenario, $35,000$ MCMC iterations were run with a burn-in period of $5,000$ iterations holding one sample of ten. To assess convergence, monitored through log posterior, we run two chains with randomly chosen initial points.

The upcoming results contain two types of studies: one in which we test the proposed method on a simulated dataset that mimics the real dataset to be analyzed with the same number of ROIs (numerical covariates) and SNPs (categorical covariates), and in the second study we increase the number of ROIs and SNPs to verify the algorithm's performance for a higher dimensional data. The reported results applies the method for jointly selecting ROIs and SNPs. Furthermore, Section 1 in the supplementary material contains more results on simulated data where we select ROIs and SNPs separately.   

We also use the posterior probability of each model to compare DDRJ to the traditional reversible jump with uniform proposals (RJ) between models. Finally, we compare DDRJ to the LASSO and random forest (RF) in terms of MCE and AUC using a 5-fold cross-validation. All the results were run using the R software \citep{rstudio} on a \textit{Intel(R) Core(TM) i7-8565U CPU 1.80GHz} with the KW statistics being implemented using Rcpp to accelerate the proposal's computation.

\begin{table}[]
\centering
      \caption{Marginal posterior probability of inclusion, coefficients estimates with standard errors in parentheses for selected ROIs and SNPs on simulated datasets. 
      $\beta, \alpha, \delta$ are the true coefficient and $ \hat{\beta}, \hat{\alpha}, \hat{\delta} $ are their respective estimates.}
   \resizebox{\textwidth}{!}{
    {
\begin{tabular}{c c c c c c c c c}
\hline
& Covariate &  mppi& $\hat{\beta}$ &$\beta$ & $\hat{\alpha}$ &$\alpha$  & $\hat{\delta}$& $\delta$    \\ \hline
& Intercept & 1.000 &1.289 (0.280) &1.000 & -& -&- &- \\
& ROI 1& 1.000 & 1.215 (0.238)& 1.300& -& -&- &- \\
$n=210$& ROI 3&0.999 &1.669 (0.284) & 1.500& -& -& -&- \\
$g=116$& ROI 115 &0.999 & 1.490 (0.292)&1.000 & -&- & -&- \\
$m=81$& SNP 1 &0.999 &- &- &1.401(0.301) & 1.300& -2.614 (0.566)&-1.200 \\
& SNP 2 &0.999 &- &- &-1.105 (0.243)  &-1.000 &-1.186 (0.513) &-1.000 \\
& SNP 3 &0.999 &- &- &1.871 (0.336) & 1.500&-0.840 (0.420) &-1.300 \\
& SNP 4 &0.999 &- &- & 1.184 (0.230)& 1.000& -2.439 (0.720) &-2.000 \\
\hline
& Intercept & 1.000& 1.436 (0.377)&1.000 & -& -&- &- \\
& ROI 1& 0.998 &1.303 (0.285) &1.300 & -& -&- &- \\
$n=300$& ROI 3& 0.998&1.886 (0.406) &1.500 & -& -& -&- \\
$g=300$& ROI 299 &0.998 &1.323 (0.313) &1.000 & -&- & -&- \\
$m=300$& SNP 1 &0.999  &- &- &1.492 (0.351) &1.300 & -1.605 (0.567) &-1.200 \\
& SNP 2 &0.878 &- &- &-0.964 (0.421) &-1.000 & -1.362 (0.660) & -1.000\\
& SNP 3 & 0.999 &- &- &1.820 (0.388) &1.500 & -1.579 (0.485)& -1.300\\
& SNP 4 & 0.999 &- &- &1.441 (0.323) &1.000 &-3.063 (0.670) & -2.000\\
\hline
& Intercept & 1.000& 1.368 (0.292)&1.000 & -& -&- &- \\
& ROI 1& 0.999 &1.716 (0.276) &1.300 & -& -&- &- \\
$n=300$& ROI 3& 0.999&1.999 (0.318) &1.500 & -& -& -&- \\
$g=500$& ROI 499 &0.998 &1.131 (0.217) &1.000 & -&- & -&- \\
$m=500$& SNP 1 &0.999  &- &- &1.343 (0.247) &1.300 & -2.480 (0.542) &-1.200 \\
& SNP 2 &0.999 &- &- &-1.101 (0.232) &-1.000 & -1.240 (0.390) & -1.000\\
&  SNP 3 & 0.999 &- &- &2.035 (0.323) &1.500 & -1.761 (0.458)& -1.300\\
& SNP 4 & 0.999 &- &- &1.379 (0.258) &1.000 &-2.834 (0.556) & -2.000\\
\hline
& Intercept & 1.000& 1.319 (0.243)&1.000 & -& -&- &- \\
& ROI 1& 0.998 &1.361 (0.214) &1.300 & -& -&- &- \\
$n=300$& ROI 3& 0.998&1.663 (0.253) &1.500 & -& -& -&- \\
$g=1000$& ROI 999 &0.998 &1.001 (0.170) &1.000 & -&- & -&- \\
$m=1000$ & SNP 1 &0.999  &- &- &1.438 (0.233) &1.300 & -1.426 (0.376) &-1.200 \\
& SNP 2 &0.878 &- &- &-1.243 (0.208) &-1.000 & -1.413 (0.386) & -1.000\\
& SNP 3 & 0.999 &- &- &1.685 (0.230) &1.500 & -2.193 (0.470)& -1.300\\
& SNP 4 & 0.999 &- &- &1.047 (0.196) &1.000 &-2.292 (0.408) & -2.000\\
\hline
\end{tabular}}
}
\label{estimates_sim_roi_snp_ddrj}
\end{table}

For the joint selection of ROIs and SNPs, the first dataset is a simulation of $g=116$ ROIs from a multivariate normal distribution with empirical  mean and covariance matrix retrieved from the  real  ROIs design matrix and we simulate $m=81$ SNPs from independent discrete distributions with probabilities retrieved from the real SNP dataset for $n=210$ individuals.  The second group of dataset contains a simulation from a standard multivariate normal and independent discrete distribution with increased number of ROIs $(300, 500, 1000)$ and SNPs $(300, 500, 1000)$, respectively. A very small number of ROIs and SNPs were chosen to have non null effects, summarized in Table \ref{estimates_sim_roi_snp_ddrj}, to maintain the proportion of healthy and diagnosed with schizophrenia. The disease status was generated using the probit model in Equation \eqref{model_geral} with prior variance set to $ \sigma_{\boldsymbol{\beta}}^2=\sigma_{\boldsymbol{\alpha}}^2=\sigma_{\boldsymbol{\delta}}^2 = 25$.

As the number of candidate variable under consideration grows $600$, $1000$, $2000$ for joint selection, we observed that a two steps procedure in which a separate pre-selection phase using a low threshold for the mppi provides better convergence. More specifically, in the first step, we separately run our method to pre-select ROIs and SNPs using a low threshold $(0.1)$ for mppi. This strategy reduces the number of covariates to approximately $10-15\%$, on average. The selected variables are then used together in the second step for joint selection and prediction. 

In summary,  DDRJ performed well in all the scenarios, selecting all the relevant variables as well as providing good estimates and small standard errors summarized in Table  \ref{estimates_sim_roi_snp_ddrj}.  Furthermore, the proposed method usually selects the true model with a higher posterior probability compared to the RJ with uniform proposals \citep{green1995reversible} as it is shown in Table  \ref{compare_rois_snps_sim_ddrj_rj}. These differences are probably due to the fact that DDRJ has been stuck for less time on wrong models since candidates are proposed in a more informative way. Finally, regarding predictive performance, in Table    \ref{prediction_simulated_rois_snps} the MCE and AUC computed from the Bayesian model averaging  show that DDRJ generally outperforms the random forest \citep{breiman1984classification} and is comparable to the LASSO \citep{tibshirani1996regression}, another well established method for variable selection.

\begin{table}[H]
    \centering
       \caption{Comparing the DDRJ and RJ using the three most visited models with their posterior probability (in parentheses) for ROIs and SNPs joint selection, where the true model column shows the true active ROIs and SNPs in the simulated model.}
  \resizebox{\textwidth}{!}{%
        {
    \begin{tabular}{c c c c}
    \hline
    & True model & DDRJ & RJ \\ \hline
   $n=210$ & ROIs (1,3,115) & ROIs (1,3,115) -- SNPs (1,2,3,4) (0.920) & (1,3,115) -- (1,2,3,4) (0.901) \\
    $m=81$   & SNPs (1,2,3,4) & (1,3,107,115) -- (1,2,3,4) (0.018)  & (1,3,7,115) -- (1,2,3,4) (0.025)    \\
   $g=116$&  & (1,3,7,115) -- (1,2,3,4) (0.013) &(1,3,49,115) -- (1,2,3,4) (0.019) \\
   \hline   
      $n=300$ & ROIs (1,3,299) & (1,3,299) -- (1,2,3,4) (0.903) & (1,3,299) -- (1,2,3,4) (0.873)    \\
    $m=300$  &  SNPs (1,2,3,4) & (1,3,75,115) -- (1,2,3,4) (0.086)  & (1,3,75,115) -- (1,2,3,4) (0.102)  \\
   $g=300$ &  & (1,3,16,115) -- (1,2,3,4) (0.006)  & (1,3,16,115) -- (1,2,3,4) (0.003)   \\
   \hline   
      $n=300$ & ROIs (1,3,499) & (1,3,499) -- (1,2,3,4) (0.834)  & (1,3,499) -- (1,2,3,4) (0.807)    \\
    $m=500$ & SNPs (1,2,3,4) & (1,3,499) -- (1,2,3,4,63) (0.113)  & (1,3,499) -- (1,2,3,4,63) (0.049)    \\
   $g=500$ & & (1,3,499) -- (1,3,4,63) (0.011) & (1,3,499) -- (1,3,4,63) (0.003) \\
   \hline   
      $n=300$ & ROIs (1,3,999) & 
      (1,3,999) -- (1,2,3,4) (0.770) & (1,3,999) -- (1,2,3,4) (0.773)     \\
    $m=1000$ & SNPs (1,2,3,4) & (1,3,528,999) -- (1,2,3,4) (0.220)  & (1,3,528,999) -- (1,2,3,4) (0.221) \\
   $g=1000$&  & (1,3,999) -- (1,3,4) (0.005)  & (1,3,999) -- (1,3,4) (0.001) \\
   \hline   
    \end{tabular} } }  
    \label{compare_rois_snps_sim_ddrj_rj}
\end{table}

\begin{table}[H]
    \centering
      \caption{Comparing the  predictive performance in terms of misclassification error (MCE) and area under the ROC curve (AUC) on simulated ROIs-SNPs dataset. In parentheses, we show the associated standard error.}
     \resizebox{\textwidth}{!}{%
          {
    \begin{tabular}{l l c c c}
    \hline 
    & & DDRJ & LASSO & RF \\ \hline 
       $n=210$, $m=81$, & MCE  & 0.193 (0.061) & 0.208 (0.026) &  0.347 (0.054)\\ 
      $g=116$& AUC & 0.880 (0.053)&  0.890 (0.023)& 0.758 (0.037)\\ \hline
      $n=300$, $m=300$, & MCE  & 0.113 (0.026) &  0.149 (0.042)& 0.302 (0.070) \\ 
        $g=300$& AUC & 0.960 (0.017) & 0.944 (0.025)& 0.791 (0.074) \\ \hline
        $n=300$, $m=500$, & MCE  & 0.156 (0.069)& 0.182 (0.047) & 0.409 (0.040) \\ 
        $g=500$ & AUC & 0.926 (0.040)& 0.899 (0.033) & 0.673 (0.044) \\ \hline
        $n=300$, $m=1000$,  & MCE  & 0.183 (0.035) & 0.136 (0.059) & 0.349 (0.028) \\ 
        $g=1000$ & AUC & 0.902 (0.029)&  0.945 (0.036)&0.743 (0.021) \\ \hline
    \end{tabular}} 
    }
    \label{prediction_simulated_rois_snps}
\end{table}

\section{MCIC data analysis}
\label{mcic}
The available dataset was collected by the MCIC \citep{chen2012multifaceted} as an effort of deeper understanding of mental disorder. It contains both imaging data on activation patterns using fMRI during a sensorimotor task and multiple SNPs allele frequencies which have previously been implicated in schizophrenia on $118$ healthy controls and $92$ individuals affected by this disorder. None of the individuals presents history of substance abuse and are free of any medical, neurological or psychiatric illnesses. Following the same approach from \cite{chekouo2016} and \cite{stingo2013integrative}, the 5-folds cross-validation with $94$ healthy controls and $74$ patients for the training set and $24$ healthy controls and $18$ patients for the validation set are used for predictive performance analysis. 

The goal of the MCIC study, a joint  effort of four research teams from Boston, Iowa, Minnesota and New Mexico, was to identify regions of interest (ROI) in the brain with discriminating activation patterns between cases and controls and relate them to a relevant set of SNPs able to explain these variations, a model selection problem clearly. The data were then preprocessed in SPM5 (\url{http://www.fil.ion.ucl.ac.uk/spm}), realigned to correct for the individuals movements, spatially normalized  to correct for anatomic variability, spatially smoothed to improve signal to noise ratio. For each of the $116$ ROIs, the activation level was summarized as the median  of the statistical parametric map values \citep{friston1994statistical}  for that region. 
The genetic information of the available dataset is given by $81$ SNPs, already known to be related to schizophrenia retrieved from the  Schizophrenia Research Forum (\url{http://www.schizophreniaforum.org/}) information. In the original dataset, the SNP information was coded as the number of minor allele for those with genotype aa, aA and AA respectively. More details of the experimental study and preprocessing can be found in \cite{chen2012multifaceted} and \cite{stingo2013integrative}. 

For each scenario, $35,000$ MCMC iterations were run with a burn-in period of $5,000$ iterations holding each sample of 10. The prior variance is set to $ \sigma_{\boldsymbol{\alpha}}^2=\sigma_{\boldsymbol{\beta}}^2=\sigma_{\boldsymbol{\delta}}^2= 25$ to ensure that the prior is not too informative but also not too vague. We ran three independent models, where two consider only ROIs or SNPS as covariates and a third model for joint selection. 

When considering ROIs as the only available covariates, the selected variables are ROIs 61 and 115 with mppi $0.837$ and $0.932$, respectively, but also suggesting more investigation on ROI 35 with mppi $0.416$ as shown in Table \ref{mcic_roi_full_coef_snp_roi_e_snp}.
ROIs 35 (left posterior cingulate region) and 61 (left inferior parietal region) were also selected by \cite{stingo2013integrative} and \cite{chekouo2016} and are known to be related to schizophrenia. In particular, ROI 115 (posterior inferior vermis--lobule IX) was a new finding that could narrow future research on lobules I to X. \cite{chekouo2016} found one more ROI 57 that has not been selected here but was present in the top 3 models. A more careful approach may be based  on this rule, including all the covariates that appear in the top 3 models to select the ROIs and consider ROIs 35, 57, 61, 96 and 115.

\begin{table}[H]
    \centering
     \caption{Marginal posterior probability of inclusion and estimates (in parentheses, we show their standard errors) for selected ROIs and SNPs on the real  dataset using either ROIs or SNPs and both of them as covariates. } 
     \resizebox{\textwidth}{!}{%
    {
    \begin{tabular}{l c c c c c}
    \hline
 Covariates &   Selected & mppi & $\hat{\beta}$ & $\hat{\alpha}$ & $\hat{\delta}$ \\ \hline
         &Intercept  & 1.000 & 0.183 (0.095)  & -& - \\
    ROIs  &   ROI 35 & 0.416 & -0.181 (0.239) & -& - \\
        & ROI 61 & 0.837 & -0.514 (0.286) & -& - \\
        &ROI 115& 0.932 & -0.607 (0.233) & -& - \\ \hline
        &    Intercept  & 1.000 & 2.511 (0.349) & - \\
    SNPs & 22 & 0.957 & -& -1.513 (0.248) & 3.842 (0.664)\\
       & 32 & 0.345 & -& 0.874 (0.482) & 0.817 (0.462)\\
       &  61 & 0.719 & -& -2.159 (0.926)&  -1.960 (0.844)\\ \hline
          &  Intercept & 1.000 & 2.945 (0.447) & - & -\\
  ROIs + SNPs    & ROI 35  & 0.291 & -0.119  (0.203) & - & - \\
      & ROI 61  & 0.794 & -0.479 (0.296) & - & - \\
      & ROI 115  & 0.968 & -0.619 (0.196) & - & - \\
      & SNP 22  & 0.955 & - & -1.602 (0.635) & 2.607 (0.592) \\ \hline     
         
    \end{tabular}} 
    }
    \label{mcic_roi_full_coef_snp_roi_e_snp}
\end{table}

Considering only the SNPs,  as shown in Table \ref{mcic_roi_full_coef_snp_roi_e_snp}, the selected variables are SNPs 22 and 61 with mppi 0.96 and 0.72, respectively. Although having a mppi 0.34 lesser than $0.5$, we also suggest SNP 32. SNP 22 (rs3737597) is located in gene DISC1 (chromosome 1), a gene  known to be strongly associated  to schizophrenia and was also found by \cite{stingo2013integrative} and \cite{chekouo2016} who also found SNPs 10 and 38 to be discriminatory.

For the joint selection of ROIs and SNPs, again ROIs 35, 61 and 115 and SNP 22 are identified as discriminatory variables with mppi 0.291, 0.794, 0.968 and 0.955, respectively. In Table \ref{mcic_roi_full_coef_snp_roi_e_snp}, we summarize the mppi, estimates and standard errors for each coefficient. Although the ROI 35 presents an mppi of less than 0.50 in the joint model, we keep it in the fitted model.  

Regarding prediction  evaluated using a 5-folds cross-validation strategy, in Table \ref{predictive_real_roi_sp_roi_e_snp} we show that DDRJ combined with Bayesian model averaging performs well in terms of predictive performance compared to the results from \cite{chekouo2016} (benchmark), LASSO, random forest  even though it is not a method focused on best prediction. 

\begin{table}[H]
    \centering
        \caption{Comparing the predictive performance on the real  dataset, using either ROIs or SNPs and both of them as covariates, in terms of misclassification error (MCE) and area under ROC curve (AUC). In parentheses, we show the associated standard error.  }
    \resizebox{\textwidth}{!}{%
    {
    \begin{tabular}{l l c c c c }
    \hline
     Covariates  &   & Benchmark & DDRJ & LASSO & RF \\ \hline
     ROIs &  MCE &  0.37 (0.02) &0.40 (0.05) & 0.38 (0.06) & 0.35 (0.05) \\ 
         & AUC & 0.66 (0.02) &0.62 (0.06)& 0.65 (0.06) & 0.68 (0.06) \\ \hline 
     SNPs & MCE &  0.45 (0.01) &0.47 (0.03) & 0.45 (0.04) & 0.44 (0.03) \\ 
         & AUC & 0.64 (0.02) &0.57 (0.02)& 0.56 (0.04) & 0.56 (0.05) \\ \hline
ROIs + SNPs &   MCE &  0.33 (0.02) &0.43 (0.02) & 0.41 (0.04) & 0.40 (0.01) \\ 
        & AUC & 0.69 (0.03) &0.67 (0.05)& 0.62 (0.04) & 0.63 (0.06) \\ \hline        
    \end{tabular}}
    }
    \label{predictive_real_roi_sp_roi_e_snp}
\end{table}

\section{Discussion}
\label{conclusao}
In this work, we have proposed a data driven reversible jump for variable selection using a  Bayesian probit model. More specifically, for identifying relevant variables that impact and regulate dichotomous traits in genetics, for which thousands of genetic, environmental and imaging information are available nowadays. The proposed method does not need the inclusion of auxiliary indicator variables for each available covariate which indicate whether it is active in the model and are updated in each MCMC iteration and the estimation of all possible models. This makes the algorithm scalable for high-dimensional data when a huge number of covariates are considered.

Our goals, selecting ROIs and SNPs and assessing predictive risk for schizophrenia based on  fMRI and SNPs information have been reached. Most ROIs 35, 57, 61, 115 and SNP 22 that we selected were in accordance with results from other authors and also known to be related to the disease, even though some new findings ROI 96 and SNPs 32 and 61 have been suggested and could be the subject of deeper research. Compared to other predictive methodologies as traditional LASSO and random forest, in terms of predictive accuracy, the DDRJ also perfoms well when predictions are done using the Bayesian model averaging, even if that is not usually the main focus.   

From a methodological perspective, we noticed that the measure (KW or Pearson correlation) used inside the DDRJ to propose the candidate model can improve or degrade the efficiency of the algorithm, as those as mainly capturing linear association.  Thus one could use some kernel based measure that accounts for non-linear relations to propose the new feature.

Regarding extensions, another direction of study would be testing other priors such as those shrinkage priors introduced earlier to improve our current methodology and evaluate the effect of the prior variance in these scenarios. As we have also mentioned, a distance matrix between ROIs is available and has not been used in this work. This information could be included either as part of the DDRJ to make better jumps, or  assume a Markov random field type of prior for ROIs and apply the DDRJ to perform variable selection and prediction for future subjects. Other extension of this work that is worth investigating is to perform clustering while selecting discriminating ROIs and SNPs, and again the DDRJ could be used to select the number of cluster and estimate parameters. 


\paragraph{\textbf{Data availability}}
The R codes and dataset used for implementing the methodologies are openly available in a public repository on Github at \url{https://github.com/hansamos/DDRJ}.
\paragraph{\textbf{Supplementary material}}
Supplementary material is available online and contains more results on selection of ROIs and SNPs separately.

\section{Author contributions statement}
D.M., D.Z., L.M., T.C. conceived the methodology.  D.M. wrote the code and conducted the experiments. D.M and D.Z wrote the manuscript. L.M and T.C analyzed and reviewed the manuscript.


\bibliographystyle{elsarticle-harv} 
\bibliography{reference}

\end{document}


\begin{frontmatter}



\title{ Supplementary material to ``Efficient Bayesian variable selection with reversible jump MCMC in imaging genetics: an application to schizophrenia"}

 \author[inst1]{Djidenou Montcho}
 \affiliation[inst1]{organization={Statistics Program, CEMSE, King Abdullah University of Science and Technology},
             city={Thuwal},
             postcode={23955600},
             country={Kingdom of Saudi Arabia}}
 \author[inst2]{Daiane A. Zuanetti}
 \affiliation[inst2]{organization={Departamento de Estatistica, Universidade Federal de Sao Carlos},
             city={Sao Carlos},
             postcode={13565905},
             state={Sao Paulo},
             country={Brazil}}
 \author[inst3]{Thierry Chekouo}
 \affiliation[inst3]{organization={Division of Biostatistics, University of Minnesota},
             city={Minneapolis},
             postcode={55455},
             state={Minnesota},
             country={USA}}

 \author[inst2]{Luis A. Milan}

\end{frontmatter}




\section{Additional simulation results} \label{simulacao_supp}
\paragraph{\textbf{Selecting ROIs (numerical covariates)}}
To mimic the real ROI dataset, we  simulate $g=116$ covariates from a multivariate normal distribution with empirical  mean and  covariance matrix retrieved from the real design matrix for $n=210$ individuals. The second group  of datasets is simulated from a  standard multivariate normal distribution  with fixed sample size $n=300$ and increased number of ROIs $( 300, 500, 1000 )$. From these covariates, we select some ROIs  with non-null effects  and their coefficients were assigned to maintain the healthy and diagnosed with schizophrenia proportion ($43.8\%$). The disease status was generated from the probit model in Equation (1) without SNPs informations and with regression coefficients summarized in Table \ref{estimates_sim_roi_ddrj}. The prior variance is set to $\sigma_{\boldsymbol{\beta}}^2 =100$ and we decide to select a ROI if its marginal posterior probability of inclusion (mppi) is greater than $0.5$.

\paragraph{\textbf{Selecting SNPs (categorical covariates)}}
Regarding the genetic dataset, we simulate $m=81$ features from independent discrete distributions with empirical probabilities retrieved from the real SNP dataset, while the second group of datasets is simulated from independent discrete distribution   with fixed sample size $n=300$ and increased number of SNPs $( 300, 500, 1000 )$. Then, we select some SNPs with non null effects and coefficients assigned to maintain the healthy and diagnosed with schizophrenia proportion.
The disease status was generated from the probit model in Equation (1) without considering ROI informations and using regression coefficients summarized in Table \ref{estimates_sim_snp_ddrj}. The prior variance is set to $\sigma_{\beta_0}^2 =\sigma_{\boldsymbol{\alpha}}^2= \sigma_{\boldsymbol{\delta}}^2= 100$.
Again, we decide to select a SNP if its mppi is greater than $0.5$.

\bigskip
In summary, DDRJ performed well in all the scenarios, selecting all the relevant variables as well as providing good estimates and small standard errors summarized in Tables \ref{estimates_sim_roi_ddrj} and  \ref{estimates_sim_snp_ddrj} for ROIs, SNPs  respectively.  Furthermore, the proposed methodology always selects the true model compared to the RJ with uniform proposals as it is shown in Tables \ref{compare_rois_sim_ddrj_rj} and \ref{compare_snps_sim_ddrj_rj}  with those differences probably due to the faster convergence of DDRJ and better mixing of DDRJ chains. Finally,  regarding predictive performance, in Tables  \ref{prediction_simulated_rois} and \ref{prediction_simulated_snps},   the MCE and AUC computed from the Bayesian model averaging  show that DDRJ generally outperforms the random forest and is comparable to the LASSO, another well established method for variable selection.

\begin{table}[H]
    \centering
    \renewcommand\thetable{A}
     \caption{Marginal posterior probability of inclusion and coefficients' estimates (in parentheses, we show their standard errors) for selected ROIs on simulated datasets. } 

     \resizebox{\textwidth}{!}{
    {
    \begin{tabular}{c c c c c}
    
     \hline
   & Selected covariate & mppi & Coef estimate & True  \\ \hline
    &  (Intercept)& 1.000& 0.794 (0.184) & 1.000  \\ 
    &  ROI 1 & 0.999& -2.020 (0.376) & -2.000  \\ 
  $n=210,\;g=116$  & ROI 3 & 0.999&-2.640 (0.526) & -2.500  \\ 
    &  ROI 115 & 0.999& 3.068 (0.496) & 3.000  \\ \hline
    
     &    (Intercept)& 1.000& 0.833 (0.156) & 1.000  \\ 
     & ROI 1 & 0.999& -0.992 (0.164) & -1.000  \\ 
    $n=300,\;g=300$ & ROI 3 & 0.999& -1.770 (0.234) & -1.500  \\ 
     & ROI 299 & 0.999& 1.968 (0.272) & 2.000  \\ \hline
    
    &   (Intercept)& 1.000& 1.202 (0.212) & 1.000  \\ 
    & ROI 1 & 0.999& -1.306 (0.248) & -1.000  \\ 
    & ROI 3 & 0.999& 0.887 (0.184) & 0.800  \\ 
$n=300$, $g=500$    & ROI 4& 0.999& -1.535 (0.233) & -1.500  \\ 
    & ROI 486 & 0.627& -0.340 (0.291) & 0.007  \\ 
    & ROI 499 & 0.999& 2.145 (0.331) & 2.000  \\ \hline
    
  & (Intercept)& 1.000& 0.957 (0.278) & 1.000  \\ 
   & ROI 1 & 0.999& 1.272 (0.330) & 1.200  \\ 
    & ROI 2 & 0.999& 0.903 (0.266) & 0.800  \\ 
$n=300,\; g=1000$    & ROI 3 & 0.999& -1.728 (0.461) & -1.500  \\ 
    & ROI 4 & 0.999& -1.206 (0.351) & -1.000  \\ 
     & ROI 1000 & 0.999& 2.840 (0.692) & 2.300  \\ 
   \hline
    \end{tabular}}
    }
    \label{estimates_sim_roi_ddrj}
\end{table}

\begin{table}[H]
    \centering
    \renewcommand\thetable{B}
      \caption{Marginal posterior probability of inclusion and estimates (in parentheses, we show their standard errors) for selected SNPs on simulated datasets.}
     \resizebox{\textwidth}{!}{
    {
    \begin{tabular}{c c c c c c c}
    \hline
& Covariate &  mppi& $\hat{\alpha}$ &$\alpha$  & $\hat{\delta}$& $\delta$    \\ \hline
 &  Intercept ($\beta_0$) &  1.000 &  1.640 (0.293) & 1.700 & -&-\\ 
 &   SNP 1 &0.999& 1.479 (0.235) & 1.300 & -0.538 (0.353)   & -1.000 \\ 
 $n=210,\; m=81$ &  SNP 2 &0.999&1.025 (0.199) & 1.000  &  -1.596 (0.409) & -1.400\\ 
  &  SNP 3 &0.998&-1.545 (0.248) & -1.500  &  -1.627 (0.431) & -1.400\\ 
  &  SNP 4  &0.999&-0.954 (0.180) & -1.200 &  -1.682 (0.437) & -2.000\\ 
   \hline
  & Intercept ($\beta_0$) &  1.000 &  2.129 (0.273) & 2.000 & -&-\\ 
  &  SNP 1 &0.999& 1.324 (0.189) & 1.300 & -1.560 (0.399)   & -1.000 \\ 
$n=300,\; m=300$  &  SNP 2 &0.999&1.320 (0.185) & 1.200  &  -1.107 (0.294) & -1.400\\ 
  &  SNP 3 &0.998&-0.956 (0.174) & -1.000  &  -1.633 (0.382) & -1.500\\ 
  &  SNP 4  &0.999&-1.662 (0.212) & -1.500 &  -1.919 (0.340) & -2.000\\  
    \hline
 &  Intercept ($\beta_0$) &  1.000 &  1.260 (0.187) & 1.300 & -&-\\ 
  &  SNP 1 &0.999& 1.135 (0.150) & 1.300 & -0.721 (0.274)   & -1.000 \\ 
 $n=300,\;m=500$ &  SNP 2 &0.998&0.933 (0.139) & 1.200  &  -1.772 (0.316) & -1.400\\ 
  &  SNP 3 &0.994&-0.912 (0.143) & -1.000  &  -1.087 (0.293) & -1.500\\ 
  &  SNP 4  &0.996&-0.414 (0.119) & -0.500 &  -1.631 (0.301) & -2.000\\ 
   \hline
 &     Intercept ($\beta_0$) &  1.000 &  1.213 (0.179) & 1.300 & -&-\\ 
 &   SNP 1 &0.998& 1.291 (0.166) & 1.300 & -1.336 (0.286)   & -1.000 \\ 
 $n=300,\; m=1000$ &  SNP 2 &0.998&1.001 (0.147) & 1.200  &  -1.393 (0.341) & -1.400\\ 
  &  SNP 3 &0.998&-0.743 (0.142) & -1.000  &  -1.390 (0.308) & -1.500\\ 
  &  SNP 4  &0.999&-0.475 (0.122) & -0.500 &  -2.092 (0.396) & -2.000\\ 
   \hline
    \end{tabular}}
    }
    \label{estimates_sim_snp_ddrj}
\end{table}

\begin{table}[H]
    \centering
    \renewcommand\thetable{C}
       \caption{Comparing the DDRJ and RJ using the three most visited models with their posterior probability (in parentheses) for ROIs selection, where the true model column shows the true active ROIs in the simulated model.}
       \resizebox{\textwidth}{!}{
    {
    \begin{tabular}{c c c c}
    \hline
    & True model & DDRJ & RJ \\ \hline
 &  &    1  3  115 (0.342) & 1  3 115 (0.304)   \\ 
$n=210,g=116$ & 1  3 115 & 1  3 70 115 (0.045)   & 1  3 70 115 (0.112)    \\
&  & 1  3 52 115 (0.039)  & 1  3 52 115 (0.042)  \\\hline   
 &   &  1 3 299 (0.341)    & 1 3 299 (0.329)  \\
$n=300, g=300$& 1 3 299 & 1  3 34 299 (0.026)  &  1 3 269 299 (0.029) \\  & & 1  3 32 299 (0.020)  & 1 3 32 299 (0.023)  \\\hline   

    &   &   1   2   3 499 (0.064)  &  1 2 3 486 499 (0.039) \\ 
$n=300,g=500$&  1   2 3   499 & 1   2   3 486  499 (0.061) & 1 2 3 129  302  393 486 499 (0.014) \\ &  & 1   2   3 177 486 499 (0.045) & 1 2 3 176 486 499 (0.011) \\\hline 
   &     & 1   2  3 4 1000 (0.083) & 1 2 3 4 1000 (0.076) \\
$n=300, g=1000$& 1 2 3 4 1000& 1   2 3   4 752 1000 (0.013) & 1   2 3  4  752 1000 (0.041)  \\ & & 1   2   3 4 353 1000 (0.01)    & 1   3 4 752 1000 (0.034) \\ \hline 
    \end{tabular} } }
    \label{compare_rois_sim_ddrj_rj}
\end{table}

\begin{table}[H]
    \centering
    \renewcommand\thetable{D}
       \caption{Comparing the DDRJ and RJ using the three most visited models with their posterior probability (in parentheses) for SNPs selection, where the true model column shows the true active SNPs in the simulated model.} 
\resizebox{\textwidth}{!}{
    {
    \begin{tabular}{c c c c}
    \hline
    & True model & DDRJ & RJ \\
    \hline
  &  &  1 2 3 4 (0.932)  & 1  2 3 4  (0.969)   \\ 
  $n=210,m=81$ & 1  2 3 4 & 1  2 3 4 75 (0.058)   & 1  2 3 4 75 (0.012)    \\ &  & 1  2 3 4 30 (0.002)  & 1  2 3 4 58 (0.005)  \\\hline   

    &   &  1  2 3 4 (0.988)    & 1  2 3 4 (0.984)\\
$n=300, m=300$& 1 2 3 4 &1  2 3 4  167 (0.004)  & 1  2 3 4 258 (0.008) \\
& & 1  2 3 4  217 (0.002)  & 1  2 3 4 17 (0.003)  \\\hline   
 &   &    1  2 3 4 (0.989)  &   1  2 3 4 (0.987)  \\
$n=300,m=500$ &  1   2 3 4 &  1  2 3 4 261(0.002) &  1  2 3 4 492 (0.001)  \\
&  &  1  2 3 4 274 (0.001)  &  1  2 3 4 417 (0.001) \\\hline 

    &     & 1   2  3 4  (0.962)  & 1 2 3 4  (0.807) \\
    $n=300, m=1000$ & 1 2 3 4 & 1   2 3   4 833 (0.006) & 1  3 (0.081)  \\ & & 1   2   3 4 990  (0.006)    & 1   2 3 (0.074) \\ \hline 
    \end{tabular}} 
   }
    \label{compare_snps_sim_ddrj_rj}
\end{table}

\begin{table}[H]
    \centering
    \renewcommand\thetable{E}
      \caption{Comparing the  predictive performance in terms of average misclassification error (MCE; in parentheses, we show its standard error) and average area under the ROC curve (AUC; in parentheses, we show its standard error) on simulated ROIs datasets. They are calculated based on these metrics observed in the test data of the 5 folds of the cross-validation scheme.}  
    {
    \begin{tabular}{l l c c c}
    \hline 
    & & DDRJ & LASSO & RF \\ \hline 
       $n=210$ & MCE  & 0.114 (0.061) & 0.137 (0.073) &  0.228 (0.064)\\ 
        $g=116$& AUC & 0.956 (0.034)&  0.944 (0.057)& 0.838 (0.054)\\ \hline
      $n=300$ & MCE  & 0.126 (0.055) &  0.129 (0.047)& 0.289 (0.035) \\ 
        $g=200$& AUC & 0.959 (0.021) & 0.944 (0.028)& 0.874 (0.018) \\ \hline
        $n=300$ & MCE  & 0.109 (0.025)& 0.149 (0.031) & 0.349 (0.016) \\ 
        $g=5000$ & AUC & 0.962 (0.020)& 0.935 (0.029) & 0.800 (0.061) \\ \hline
        $n=300$ & MCE  & 0.133 (0.042) & 0.109 (0.022) & 0.369 (0.021) \\ 
        $g=1000$& AUC & 0.942 (0.022)&  0.951 (0.015)&0.820 (0.061) \\ \hline 
    \end{tabular}}
    \label{prediction_simulated_rois}
\end{table}

\begin{table}[H]
    \centering
    \renewcommand\thetable{F}
      \caption{Comparing the  predictive performance in terms of misclassification error (MCE) and area under the ROC curve (AUC) on simulated SNPs dataset. In parentheses, we show the associated standard error.} 
    \begin{tabular}{l l c c c}
    \hline 
    & & DDRJ & LASSO & RF \\ \hline 
       $n=210$, & MCE  & 0.104 (0.031) & 0.175 (0.024) &  0.251 (0.041)\\ 
        $m=81$& AUC & 0.942 (0.020)&  0.924 (0.015)& 0.851 (0.030)\\ \hline
      $n=300$, & MCE  & 0.143 (0.049) &  0.190 (0.062)& 0.346 (0.021) \\ 
        $m=300$& AUC & 0.934 (0.033) & 0.911 (0.033)& 0.872 (0.053) \\ \hline
        $n=300$, & MCE  & 0.166 (0.065)& 0.195 (0.055) & 0.396 (0.015) \\ 
        $m=500$& AUC & 0.907 (0.004)& 0.866 (0.039) & 0.730 (0.007) \\ \hline
        $n=300$, & MCE  & 0.176 (0.060) & 0.229 (0.032) & 0.400 (0.011) \\ 
        $m=1000$ & AUC & 0.905 (0.035)&  0.864 (0.031)&0.719 (0.028) \\ \hline \end{tabular}  
    \label{prediction_simulated_snps}
\end{table}


\appendix
\section{Conditionals distribution for Gibbs sampling procedure}
\label{appendix a}

\begin{eqnarray}
\boldsymbol{\beta}|\boldsymbol{Y}^*,\boldsymbol{X,Z}, \boldsymbol{\alpha}, \boldsymbol{\delta} &\sim& N(\boldsymbol{\beta}^*, \Gamma_1),\; \boldsymbol{\beta}^* =  \Gamma_1\;[\boldsymbol{1|X}]^{T}\;\{ \boldsymbol{Y}^{*}- \boldsymbol{Z}\boldsymbol{\alpha} - [\boldsymbol{1}- |\boldsymbol{Z}|]\boldsymbol{\delta} \} , \nonumber\\  \Gamma_1 &=& \left\{\frac{1}{\sigma_{\boldsymbol{\beta}}^{2}}\boldsymbol{I}_{P+1} + [\boldsymbol{1}|\boldsymbol{X}]^{T}[\boldsymbol{1}|\boldsymbol{X}] \right\}^{-1}; 
\end{eqnarray}
\begin{eqnarray}
\boldsymbol{\alpha}|\boldsymbol{Y}^*,\boldsymbol{X,Z}, \boldsymbol{\beta}, \boldsymbol{\delta} &\sim& N(\boldsymbol{\alpha}^*, \Gamma_2),   \;
 \boldsymbol{\alpha}^* = \Gamma_{2}\;\boldsymbol{Z}^{T}\;\{ \boldsymbol{Y}^{*}- [\boldsymbol{1}|\boldsymbol{X}]\boldsymbol{\beta} - [\boldsymbol{1}- |\boldsymbol{Z}|]\boldsymbol{\delta} \}   , \nonumber \\ 
 \Gamma_2 &=&  \left\{\frac{1}{\sigma_{\boldsymbol{\alpha}}^{2}}\mathbf{I}_{K} + \boldsymbol{Z}^{T}\boldsymbol{Z} \right\}^{-1};  
 \end{eqnarray}
 
\begin{eqnarray}
\boldsymbol{\delta}|\boldsymbol{Y}^*,\boldsymbol{X,Z}, \boldsymbol{\beta}, \boldsymbol{\alpha} &\sim& N(\boldsymbol{\delta}^*, \Gamma_3),\; \boldsymbol{\delta}^* = \Gamma_3\boldsymbol{[1-|Z|]}^{T}\{ \boldsymbol{Y}^{*}- [\boldsymbol{1|X}]\boldsymbol{\beta} - \boldsymbol{Z}\boldsymbol{\alpha} \},  \nonumber \\
 \Gamma_3 &=& \left\{\frac{1}{\sigma_{\boldsymbol{\delta}}^{2}}\mathbf{I}_{K} + [\boldsymbol{1-|Z|}]^{T}[\boldsymbol{1-|Z|}] \right\}^{-1};  
 \end{eqnarray}
 
 \begin{eqnarray}
Y^{*}_{i}|\boldsymbol{\theta}, Y_{i}=y_i, \boldsymbol{X}_{i},\boldsymbol{Z}_{i}  \sim \begin{cases}
    \text{Nt}([1|\boldsymbol{X}_{i}]\boldsymbol{\beta} + \boldsymbol{Z}_{i}\boldsymbol{\alpha} + (1-|\boldsymbol{Z}_{i}|)\boldsymbol{\delta}, 1 ,\text{left}=0), y_i = 1 \\
    \text{Nt}([1|\boldsymbol{X}_{i}]\boldsymbol{\beta} + \boldsymbol{Z}_{i}\boldsymbol{\alpha} + (1-|\boldsymbol{Z}_{i}|)\boldsymbol{\delta}, 1 ,\text{right}=0), y_i = 0 
\end{cases}
\end{eqnarray}
with $ \mathbf{I}_{N}$ being the identity matrix of size $N$, $Nt$ being the truncated Normal Distribution. $[\boldsymbol{1|X}]$ is  a matrix of dimension $n\times(P+1)$ having  ones (1) in the first column, $\boldsymbol{X}_i$ and $\boldsymbol{Z}_i$ are the $ith$ row of the design matrices $\boldsymbol{X}$ and $\boldsymbol{Z}$ respectively, where the columns correspond to the selected features.

\section{Complete algorithm for the RJMCMC}
\label{appendix b}
\begin{algorithm}[H]
\tiny
\caption{DDRJ algorithm for covariates selection.}
\begin{algorithmic}
\State Input $P = K = 0$ to start without ROIs or SNPs in the model.
\State Sample $\boldsymbol{Y}^*$ from the truncated normal. 
\For{$l = 1$ to $L$}
    \State Choose a jump into either the ROIs  or SNPs  space.
    \If{Jump is into ROIs space}
        \State Choose either a birth or death move.
        \If{Birth move is chosen}
            \State Select a ROI to include using $p_{bj}$.
            \State Sample candidate  $\boldsymbol{\theta}^b$ from its full conditional. 
            \State Accept proposal with probability $\psi^b$ 
            \If{Accepted}
                \State Update model size: $P^{(l)} = P^{(l-1)} + 1$, $K^{(l)} = K^{(l-1)}$.
                \State Update parameters to $\boldsymbol{\theta}^b$ and $\boldsymbol{Y}^*$ from their full conditional.  
            \Else
                \State Retain previous model size and parameters.
            \EndIf
        \ElsIf{Death move is chosen}
            \State Select a ROI to exclude using $p_{dj}$.
            \State Sample parameters and evaluate acceptance using $\psi^d$ 
            \If{Accepted}
                \State Update model size: $P^{(l)} = P^{(l-1)} - 1$, $K^{(l)} = K^{(l-1)}$.
            \Else
                \State Retain previous model size and parameters.
            \EndIf
        \EndIf
    \ElsIf{Jump is into SNPs space}
        \State Choose either a birth or death move.
        \If{Birth move is chosen}
            \State Select an SNP to include using $p_{bk}$.
            \State Update parameters and evaluate acceptance.
            \If{Accepted}
                \State Update model size: $P^{(l)} = P^{(l-1)}$, $K^{(l)} = K^{(l-1)} + 1$.
            \Else
                \State Retain previous model size and parameters.
            \EndIf
        \ElsIf{Death move is chosen}
            \State Select an SNP to exclude using $p_{dk}$.
            \State Sample parameters and evaluate acceptance.
            \If{Accepted}
                \State Update model size: $P^{(l)} = P^{(l-1)}$, $K^{(l)} = K^{(l-1)} - 1$.
            \Else
                \State Retain previous model size and parameters.
            \EndIf
        \EndIf
    \EndIf
\EndFor
\end{algorithmic}
\end{algorithm}